\documentclass{optica-article}

\journal{opticajournal} 

\articletype{Research Article}

\usepackage{lineno}
\usepackage{subcaption}
\usepackage{physics}
\usepackage{multirow}

\begin{document}

\title{Low-Loss Polarization-Maintaining Optical Router for Photonic Quantum Information Processing}

\author{Pengfei Wang,\authormark{1,2} Soyoung Baek,\authormark{1} Keiichi Edamatsu,\authormark{1} and Fumihiro Kaneda\authormark{1,3,4,*}}

\address{\authormark{1}Research Institute of Electrical Commnuication, Tohoku University, Sendai 980-8577, Japan\\
\authormark{2}Graduate School of Engineering, Tohoku Univeristy, Sendai 980-8579, Japan\\
\authormark{3}Graduate School of Science, Tohoku Univeristy, Sendai 980-8578, Japan\\
\authormark{4}Precursory Research for Embryonic Science and Technology (PRESTO), Japan Science and Technology Agency (JST), Kawaguchi 332-0012, Japan
}

\email{\authormark{*}fumihiro.kaneda.b6@tohoku.ac.jp} 


\begin{abstract*} 
  In photonic quantum applications, optical routers are required to handle single photons with low loss, high speed, and preservation of their quantum states. Single-photon routing with maintained polarization states is particularly important for utilizing them as qubits. Here, we demonstrate a polarization-maintaining electro-optic router compatible with single photons. Our custom electro-optic modulator is embedded in a configuration of a Mach-Zehnder interferometer, where each optical component achieves polarization-maintaining operation. We observe the performance of the router with 2-4\% loss, 20 dB switching extinction ratio, 2.9 ns rise time, and $>$ 99\% polarization process fidelity to an ideal identity operation. 
\end{abstract*}

\section{Introduction}
Optical switching and optical routing are key technologies in modern optical communication networks as well as photonic quantum information and communication applications. In photonic quantum applications, an optical router is required to direct single photons into a desired optical channel \cite{rupertursin}. Such a single-photon router is a fundamental component of quantum channel multiplexers \cite{LeeNPJQ75_2022}, all-optical quantum memories \cite{kanedaoptica, kanedaSA}, and quantum repeaters \cite{repeater1, repeater2} to realize important applications including large-scale quantum computing, long-distance quantum communication, and photonic quantum-state synthesis \cite{MccuskerPRL2009}. 

Optical routers compatible with single photons need to meet three essential requirements: first, they need to be operated with low loss, since a lost single photon cannot be restored, unlike classical states of light. The loss of single photons is a significant source of error in photonic quantum computing\cite{oqc}. The second requirement is the preservation of a single photon’s quantum state except for the spatial degree of freedom. This is crucial for high-quality multi-photon interference \cite{HOM_PRL1987} in photonic quantum gate operations \cite{KLMNature2001}. In particular, the polarization-maintaining operation is important since a single-photon polarization state is widely used as a qubit. The third requirement is high-speed operation, necessary for speeding up and scaling up general quantum applications.  

Several experiments have demonstrated the routing of single photons; however, they lack one or more of the above requirements. For example, all-optical approaches using nonlinear Kerr effect\cite{ultrafast,THz,THzperspective} in an optical fiber cable have achieved a \(\sim\) THz switching bandwidth, but with polarization dependence and \(>\) 20\% loss. Single-photon switching based on intra-cavity difference-frequency generation\cite{cavity} also has a large loss due to unwanted frequency conversion and intra-cavity loss. Routers based on electro-optic modulators (EOM) with bulk optics\cite{kanedaSA} and waveguide\cite{waveguideEOM} structures have demonstrated low-loss (\(\sim\) 1\%) and high-speed switching (\(>\) 10 GHz), respectively. However, individual electro-optic (EO) crystals have intrinsic and electro-optical birefringence, i.e., polarization dependence. A polarization-maintaining EO router has been demonstrated by introducing additional birefringence compensator crystals\cite{antonZ}, which, however, can also accompany loss and beam wavefront distortion degrading switching extinction ratio.

In this work, we propose and experimentally demonstrate a low-loss, polarization-maintaining EO router compatible with single photons. Our scheme is based on the polarization-maintaining operation of all optical components including custom birefringence-compensated EO crystals. This enables the construction of a router with a Mach-Zehnder interferometer (MZI) configuration without polarization compensators. Our experiment with classical laser pulses demonstrates 2-4\% loss, 20 dB switching extinction ratio (SER), and \(>\) 99\% process fidelity in the polarization degree of freedom. The measured routing rise time is as fast as a standard bulk optic EOM (2.9 ns). The EO Pockels effect employed in our experiment essentially does not produce noise photons, as demonstrated in previous quantum optics experiments.\cite{kanedaSA,JWP,silberhorn,xanadu,furusawa} Therefore, our router is expected to perform similarly for single-photon states having similar spectral and spatial modes to the classical light pulses used in our experiment. We expect that our demonstration is an important step toward advanced polarization-encoded photonic quantum information technologies that rely on photon-polarization qubits and their entanglement.

\section{Methods}
Figure 1 (a) illustrates our optical setup of the polarization-maintaining router based on an MZI with an embedded custom birefringence-compensated EOM. The EOM modulates the phase difference between two interferometer arms by applying an electric field to EO crystals, enabling us to select an output optical path.
\begin{figure}[ht]
    \centering
    \includegraphics[width=\columnwidth]{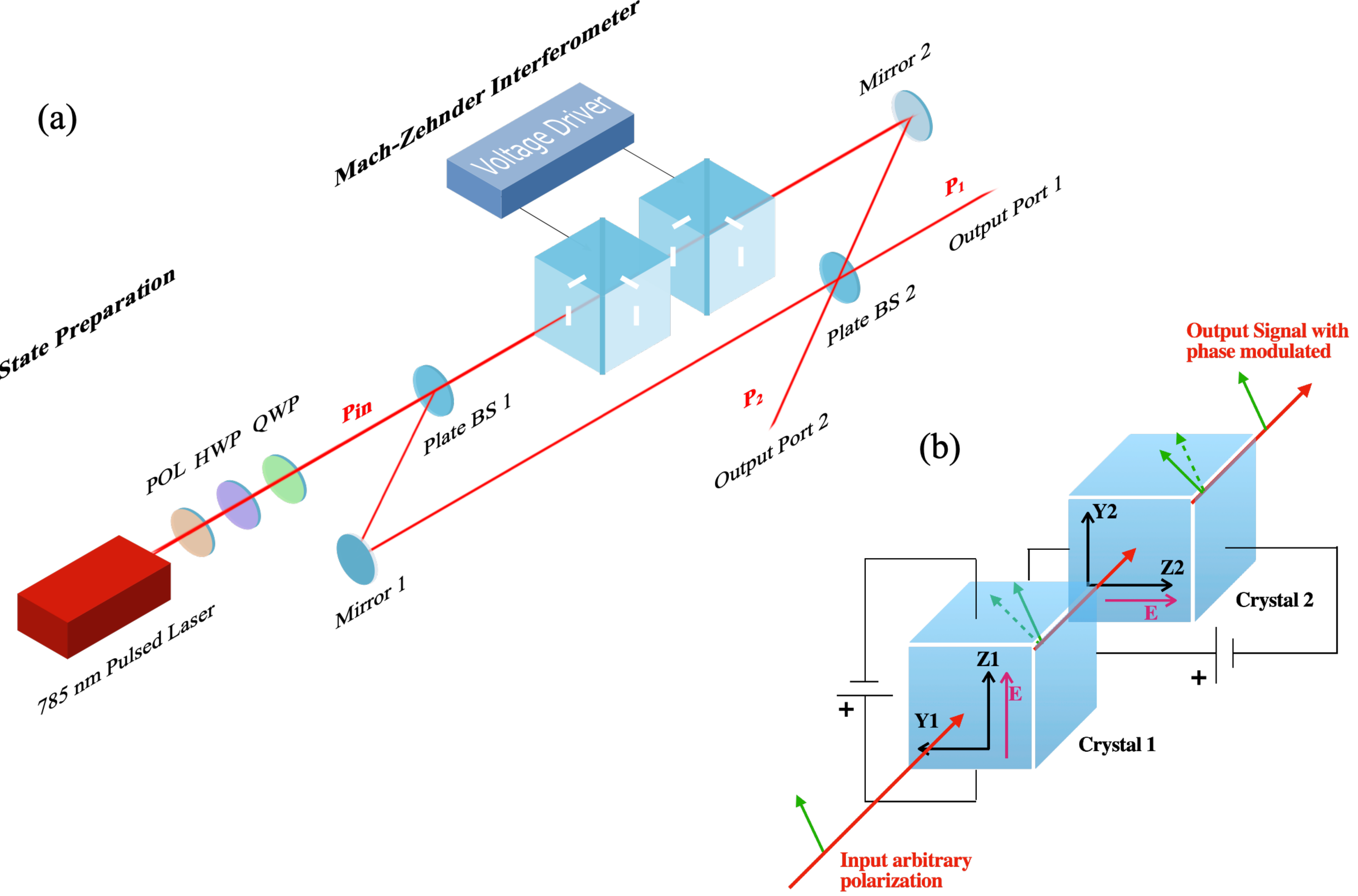}
    \caption{(a) Schematic diagram of our polarization-maintaining electro-optic (EO) router. POL, polarizer; HWP, half-wave plate; QWP, quarter-wave plate; BS, non polarizing beam splitter. \(P_{\mathrm{in}}\), \(P_1\), and \(P_2\) denote optical power at the input port and the output ports 1 and 2. (b) Polarization-maintaining EOM. Two rubidium titanyl phosphate (RTP) crystals with orthogonally oriented crystallographic axes are placed in series. Electric fields following the crystallographic \(Z\) axes are applied to two crystals. The birefringence of the two crystals is compensated for each other, enabling an identical phase shift to arbitrary input polarization states. An angle of incidence to the other optics component is nearly normal (\(\sim5^{\circ}\)), so they are operated independently of polarization.}
    \label{setup}
\end{figure}
The polarization-maintaining routing is achieved by \textit{all} constituent optical components maintaining polarization states of light: Our EOM contains two cross-axis aligned rubidium titanyl phosphate (RTP) crystals (Crystal 1 and Crystal 2) with an applied electric field along the crystallographic \(Z\) axis direction, as shown in Fig. 1 (b). This configuration resembles a thermally compensated Pockels cell, i.e., an EO polarization switch, but our EOM has the opposite polarity of an applied field to one EO crystal compared to the one in a Pockels cell.
The operation of the EOM can be described by decomposing the phase shift of horizontal (\(H\)) and vertical (\(V\)) polarizations that are parallel or perpendicular to the crystallographic \(X\) and \(Y\) directions. The phase shift $\phi_{kj}$ of the $k$ ($= H, V$) polarized light given by Crystal $j$ ($= 1, 2$) can be calculated as\cite{yariv}
\begin{equation}
\phi_{V1}=\frac{2\pi}{\lambda}l_1(n_z-\frac{1}{2}r_{33}n_z^3\frac{U_1}{d_1}),~\phi_{V2}=\frac{2\pi}{\lambda}l_2(n_y-\frac{1}{2}r_{23}n_y^3\frac{U_2}{d_2}),
\end{equation}
\begin{equation}
\phi_{H1}=\frac{2\pi}{\lambda}l_1(n_y-\frac{1}{2}r_{23}n_y^3\frac{U_1}{d_1}),~\phi_{H2}=\frac{2\pi}{\lambda}l_2(n_z-\frac{1}{2}r_{33}n_z^3\frac{U_2}{d_2}),
\end{equation}
where \(\lambda\) denotes the wavelength of the incident light; \(l_j\) and \(d_j\) are the length and width of the RTP crystals; \(n_y\) (\(n_z\)) and \(r_{23}\) (\(r_{33}\)) are the intrinsic refractive indices and EO coefficient for crystallographic \(Y\) (\(Z\)) direction, respectively; \(U_j\) is the voltage applied to the crystal. For identical applied fields to the two crystals, i.e., \(U_1/d_1 = U_2/d_2\), one can achieve the same phase shift \(\phi_{V1}+\phi_{V2}=\phi_{H1}+\phi_{H2}\) for orthogonal polarization states. Thus, the static and EO birefringence of the two crystals is compensated for each other, enabling a polarization-maintaining phase shift according to the applied electric field. We note that our scheme is generally applicable to a pair of identical EO crystals that is often fabricated for off-the-shelf thermally compensated Pockels cells, and it applies even if the crystals have different widths and/or lengths, as long as the applied voltage to the crystals is calibrated accordingly. The other optical components of the router also preserve polarization states of light: in our squished configuration of the MZI, all optics components are operated with a nearly normal angle of incidence (AOI, \(5^{\circ}\)) to mitigate the polarization dependence (on S- and P-polarizations) in transmissivity, reflectivity, and phase shift\cite{AOI1,AOI2}. The interferometer arm without EOM is 16 mm longer than the other to compensate for the intrinsic group delay (relative to air) given by the EOM (\(l_1= l_2 = 10\) mm).

\section{Results and Discussion}
We constructed a polarization-maintaining router based on the scheme described above, using optical components with anti-reflection/high-reflection coatings optimized at the wavelength of \(\lambda = 785\) nm. In this proof-of-concept experiment, we characterized our router via classical photodetection measurements of optical pulses from a diode laser at 785 nm with a bandwidth of 2 nm. The diode laser beam is coupled to a single-mode optical fiber and then collimated by an aspheric lens (with a focal length of 7.5 mm) so that an input optical pulse to the router has a Gaussian spatial mode. Our measurement predicts the router's performance on single-photon states having similar bandwidth and spatial mode to the classical laser pulses since our router utilizes the EO effect which is essentially independent of input optical power as shown in Eq. (1, 2) and produces no noise photons, as has been demonstrated in many quantum optics experiments \cite{kanedaSA,JWP,silberhorn,xanadu,furusawa}. A set of a polarizer (POL), a half-wave plate (HWP), and a quarter-wave plate (QWP) is used to prepare an arbitrary input polarization state. The router is operated at a 10 kHz repetition rate, which is limited by our high-voltage EOM driver; we expect that a \(>\) 1 MHz repetition rate is available with our EOM that is compatible with off-the-shelf high-voltage drivers designed for standard Pockels cells (used for polarization switching)\cite{kanedaSA,antonZ}. Our MZI is stable for several minutes which is limited by the environmental disturbance. Nevertheless, the stability can be improved by miniaturizing the setup and/or utilizing matured active phase stabilization techniques to be applied for advanced quantum optics experiments. In our experiment performed at the low repetition rate (10 kHz), the phase variation due to the temperature instability of the EOM is not observed. The temperature instability of the EOM that may occur at a higher repetition rate can be solved by active temperature stabilization techniques such as thermoelectric and water-cooling methods. A thermal compensation configuration can also be constructed by placing a polarization-maintaining EOM on each arm of the MZI and applying opposite voltages to them. This push-pull operation also reduces the necessary voltage for switching an output port by half.
\begin{figure}[h]
\centering
    \includegraphics[width=\linewidth]{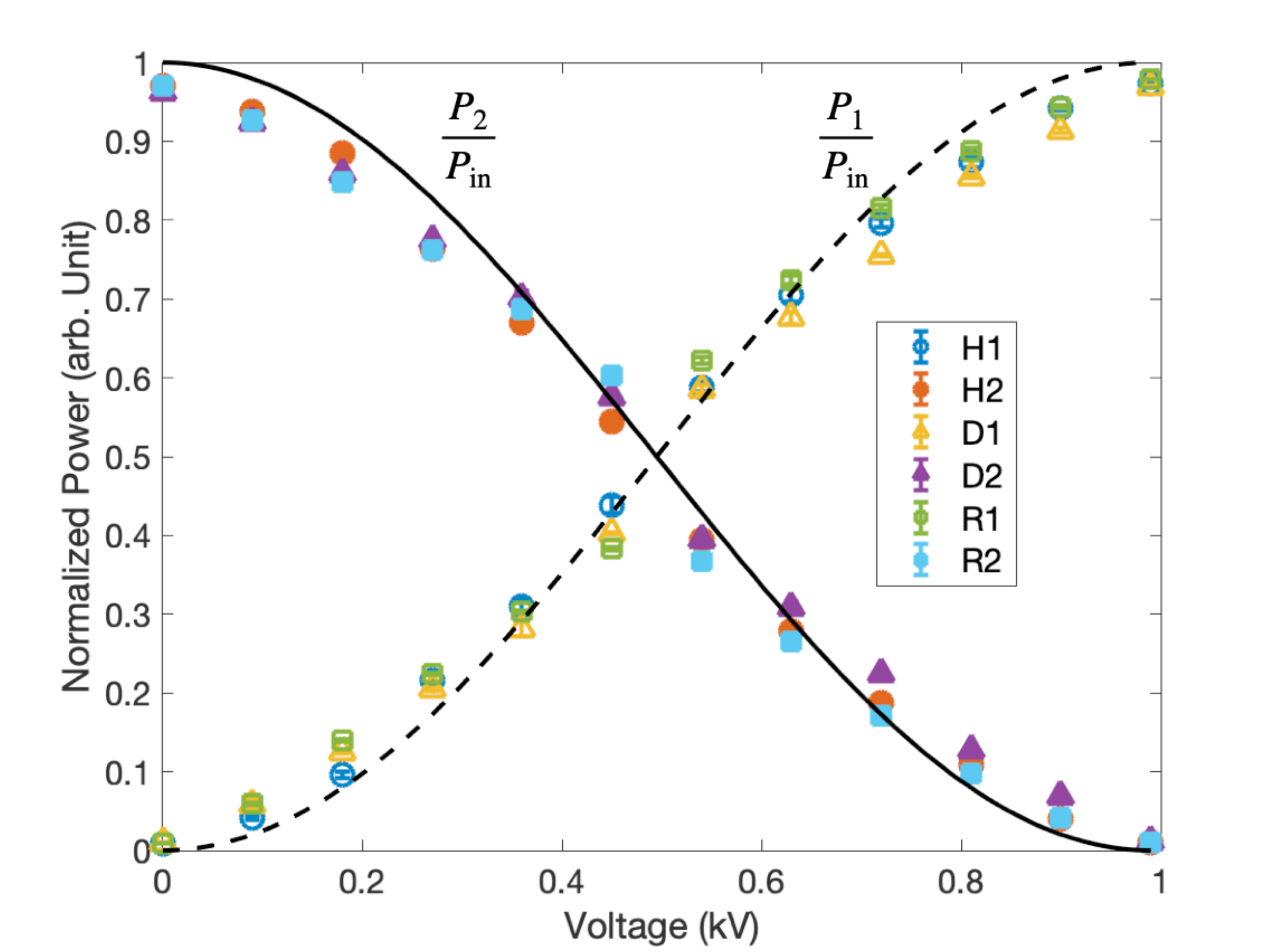}
\caption{Measured output power versus applied voltage to the EOM for horizontal (\(H\), circles), diagonal (\(D\), triangles), and right-circular (\(R\), squares) input polarization states. The blank and solid symbols respectively show the measured power at the output ports 1 and 2; For example, the result with horizontal input and the output port 1 (\(H1\)) is shown as blank circles. The solid and dashed lines are the best sinusoidal fits of the experimental data. The error bar indicates the standard deviation of 100 independent measurements. }
\label{switchingcurve}
\end{figure}

We measured optical power at the input port (\(P_{\textrm{in}}\)) and the output ports 1 and 2 (\(P_1\) and \(P_2\)) for different input polarization states and the applied voltage (\(U = U_1 = U_2\)) to the EOM. Figure \ref{switchingcurve} shows the normalized output optical power (\(P_1/P_{\textrm{in}}\) and \(P_2/P_{\textrm{in}}\)) versus \(U\). The output power for horizontal (\(H\)), diagonal (\(D\)), and right-circular (\(R\)) polarized light changes according to the applied voltage, demonstrating the full tunability of the splitting ratio between output ports 1 and 2. We observed high interference visibility (98\%) without spatial filtering. This indicates that the spatial mode of the input light is highly preserved in the router constructed by a minimum number of optics. With the measured input and output optical powers we also evaluated the SER and loss, as shown in Table \ref{SER}. The loss and SER for the output port 1 are obtained by \(1-P_1/P_{\mathrm{in}}\) and \(P_1/P_2\)  for \(U_1=U_2 = 1.0\) kV. Those for the output port 2 are obtained similarly for \(U_1=U_2 = 0\) kV. Here, the loss at each output port includes the insertion loss (\(1-P_1/P_{\mathrm{in}}-P_2/P_{\mathrm{in}}\)) and the leakage to undesired output port (e.g., \(P_2/P_{\mathrm{in}}\) for the loss at the output port 1) caused by imperfect interference visibility (98\%). For four input polarization states, the router has a loss of 2-4\% and a SER of 20 dB. The slight difference in SER and loss for the output ports 1 and 2 are caused by the imbalanced transmission and reflection (49\% versus 51\%) of the beam splitters.
\begin{table}[h]
\centering
\caption{Measured SER and loss for both output ports with $H,V,D,R$ input polarization states.}
\begin{tabular}{ l c c c c c  } 
\hline
& Input polarization & \textit{H} & \textit{V} & \textit{D} & \textit{R} \\
\hline
\multirow{2}{3cm}{Switching extinction ratio (dB)} & Output port 1 & 19.9(1) & 19.8(1) & 19.4(1) & 19.9(1) \\ 
& Output port 2 & 20.5(2) & 20.5(3) & 20.0(1) & 20.5(1) \\ 
\hline
Insertion Loss (\%) & &1.9(2)& 1.4(4)& 1.9(3)&1.3(3) \\
\hline
\multirow{2}{3cm}{Loss (\%)} & Output port 1 & 2.74(4) & 2.12(4) & 3.01(10) & 2.10(5) \\ 
& Output port 2 & 2.90(10) & 2.92(12) & 3.68(9) & 2.90(8) \\
\hline
\end{tabular}
\label{SER}
\end{table}

\begin{figure}[h]
\centering
    \includegraphics[width=\linewidth]{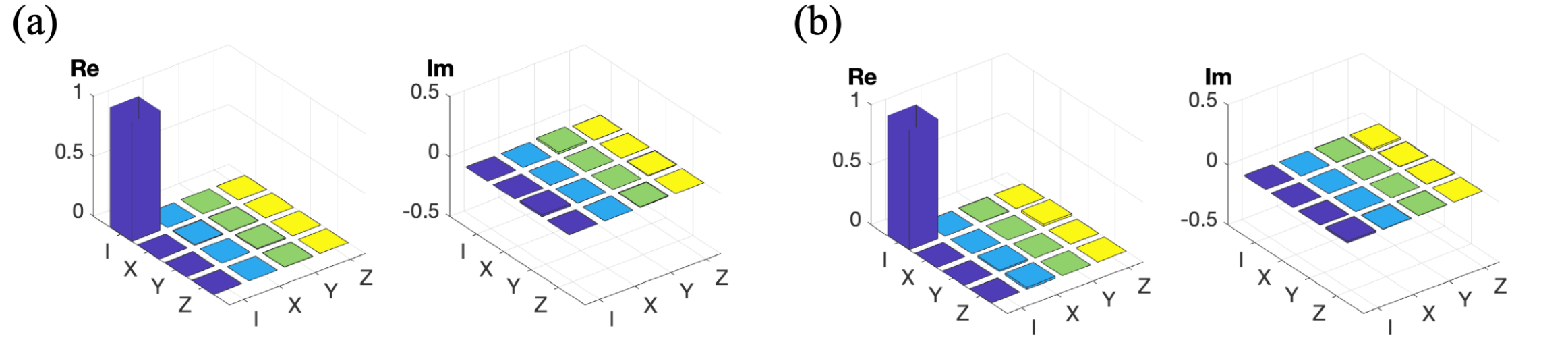}
\caption{Reconstructed process matrix \(\chi_\mathrm{e}\) in Pauli basis for (a) the output port 1 and (b) the output ports 2.The process fidelities to the ideal operation (\(\chi_{II} =1\) and other elements are zero) are 0.9948(9) and 0.9955(17) for the output ports 1 and 2, respectively.}
\label{process}
\end{figure}

To characterize the router’s polarization-maintaining capability, we performed quantum process tomography\cite{process} of the routing process, which determines the (unwanted) operation of the router applied to an input polarization density matrix \(\rho_{\mathrm{in}}\). 
The polarization process of the router is decomposed into the Pauli basis \(\sigma_i, (i= I, X, Y, Z)\) and characterized as a process matrix $\chi$, where the matrix elements \(\chi_{ij}\) are given by:
\begin{equation}
\rho_{\mathrm{out}}=\sum_{i,j}\chi_{ij}\sigma_i\rho_{\mathrm{in}}\sigma_j^{\dag}, \\
\end{equation}
where the \(\rho_{\mathrm{out}}\) denotes the output polarization density matrix.
We employed the maximum likelihood method \cite{tomography} to reconstruct the experimental process matrix \(\chi_\mathrm{e}\) by the measurement of output polarization states (via quantum state tomography) for six input eigenstates (\(H,V,D,A,R\) and \(L\) polarization states) of the three Pauli operators. Figure \ref{process} shows the reconstructed process matrix \(\chi_{\mathrm{e}}\) for two output ports. The operation of the router is close to an ideal process matrix \(\chi_\mathrm{i}\): \(\chi_{II}= 1\) and other elements are 0, with the fidelity \(F= [\Tr(\sqrt{\sqrt{\chi_{\mathrm{e}}} \chi_\mathrm{i}\sqrt{\chi_{\mathrm{e}}}})]^2\) of 
0.9948(9) and 0.9955(17) for the output ports 1 and 2, respectively (the uncertainties are obtained by the standard deviations of ten independent tomographic measurement datasets). Thus, the router demonstrates the capability of directing input light to an arbitrary output port with highly maintaining its polarization state. Note that in the output port 2 the phase of the horizontal polarization is shifted by \(\pi\) relative to the vertical polarization because of the odd number of reflections; the phase shift can be easily corrected by an additional mirror reflection. In our experiment, we used the inverted coordinate of the horizontal polarization at the output port 2, to estimate \(\chi_{\textrm{e}}\).

\begin{figure}[h]
\centering
    \includegraphics[width=\linewidth]{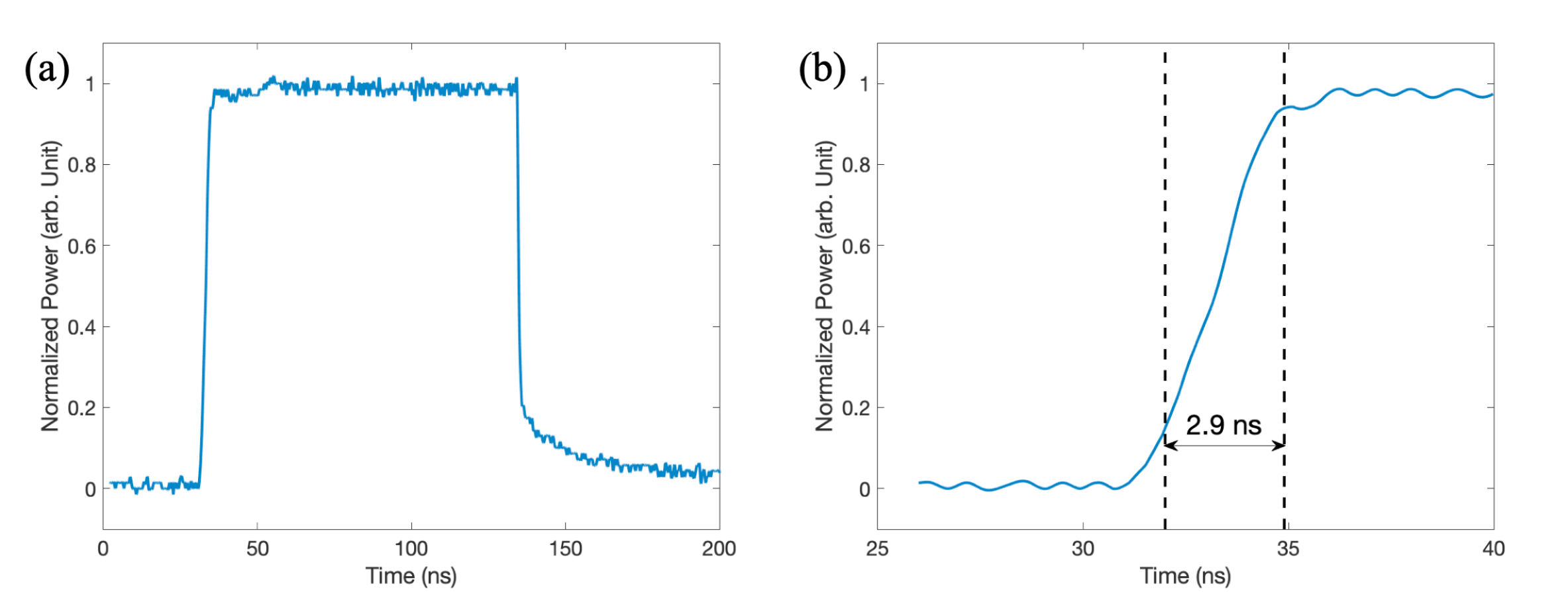}
\caption{(a) Measured switching response of our router. The operation period is 100 ns. (b) The rising edge of the switching window. The observed rise time (10\% to 90\% of signal intensity transition time) is 2.9 ns.}
\label{speed}
\end{figure}

Finally, we characterized the routing rise/fall time (10\% to 90\% of signal intensity transition time) of our router using a fast Si photodetector (rise/fall time \(<0.5\) ns) placed at the output port 1. Figure \ref{speed} shows the measured output power as a function of time with the EOM applied a half-wave voltage for 100 ns. We observed a rise time of 2.9 ns, which is as fast as a standard Pockels cell. The fall time from 90\% to 20\% is 2.2 ns, whereas 15.6 ns from 90\% to 10\% is not as fast as the rise time. This may be due to the slow recovery process in our high-voltage driver.

\begin{table}[h]
\centering
\caption{Comparison of performances of polarization-maintaining routers. In Ref. 16, the loss includes fiber coupling loss, and the polarization state fidelity for an arbitrary polarization state, instead of process fidelity, is demonstrated. }
\begin{tabular}{ p{2cm} p{2cm} p{1cm} p{1.7cm} p{2cm} p{1.7cm} } 
\hline
References& Loss (\%) & SER& Repetition&Rise time (ns)& Process \\
          &          & (dB) &rate (MHz)&               &fidelity (\%)\\
\hline
(16) & 75 & 16 & 2.5 & 5.6 & 98 \\ 
(26) & 10  & 14.8 & 0.6 & 6 & 94.7 \\ 
(27)& 86.7 \& 83.8 & 19.21 & \(10^3\) & 10 & - \\
\hline
This work& 2-4&20 & \(10^{-2}\) & 2.9 & $> 99$ \\
\hline
Possible improvements& $< 1$ &$> 20$ & $> 1$ & 2.9 & $> 99$ \\
\hline
\end{tabular}
\label{comparison}
\end{table}
Table \ref{comparison} shows the performance of our router in comparison with previous works\cite{antonZ,interface,sagnac}. Our router has outperformed in loss, SER, and polarization process fidelity. Moreover, our router has possible improvements with feasible existing technologies. The absorption loss in the EOM (\(\sim\) 2\%) can be mitigated at a longer wavelength; for example, at a telecommunication band, EOM with RTP crystals has less than 1\% insertion loss\cite{kanedaSA}. The \(\sim\) 20 dB SER is currently limited by the imperfect (98\%) interference visibility, which also introduces \(\sim\) 1\% loss at the desired output port. This may be improved by high-precision, high-stability optics and mechanical components. Although we used optical pulses with a 2-nm bandwidth for characterizing the performance of our router, the acceptance bandwidth for < 1\% degradation of the interference visibility is estimated to be 30 nm, with the group velocity dispersion (302 \(\mathrm{fs}^2/\mathrm{mm}\)) of the RTP crystals taken into account.\cite{n_RTP}. The EOM is compatible with a high-repetition-rate (\(>\) 1 MHz) driver designed for thermally compensated Pockels cells. The repetition rate can also be improved by using multiple pieces of our custom EOMs in the MZI. With the feasible improvement outlined above, the low-loss, high-speed routing of single-photon polarization qubits is possible for scaling up future photonic quantum information processing.

\section{Conclusion}
We have experimentally demonstrated a low-loss and polarization-maintaining EO router applicable to photonic quantum information experiments with a 2-4\% loss, a 20 dB SER, a 2.9 ns rise time, and a \(>\) 99\% polarization process fidelity. 
The performance is achieved by the polarization-maintaining operation of all constituent optical components including our custom EOM without polarization compensator optics. After the feasible improvements, our router can be immediately applicable to all-optical storage of photons for high-efficiency synthesis and measurement of polarization-encoded photonic quantum states \cite{migdall,pittman,multiphoton,multiplexedentangle}. Also, the splitting ratio of the router is fully tunable, enabling the usage of universal quantum gates in quantum computing\cite{KLMNature2001,oneway} and quantum metrology\cite{metrology}. We anticipate that our demonstrated scheme will be an important building block in a wide range of photonic quantum information technologies.

\begin{backmatter}
\bmsection{Acknowledgments}
This work is supported by JSPS KAKENHI Grant Number JP21K18902 and JP22H01965, MEXT Quantum Leap Flagship Program (MEXT Q-LEAP) Grant Number JPMXS0118067581, and JST PRESTO (JPMJPR2106).
\end{backmatter}

\begin{thebibliography}{1}
 \newcommand{\enquote}[1]{``#1''}
 
\bibitem{rupertursin}
 S. Wengerowsky, S. Joshi, F. Steinlechner, H. H{\"u}bel, and R. Ursin,
   \enquote{An entanglement-based wavelength-multiplexed quantum communication network,}
   {\protect\JournalTitle{Nature}} \textbf{564}, 225--228 (2018).

\bibitem{LeeNPJQ75_2022}
    Y. Lee, E. Bersin, A. Dahlberg, S. Wehner, and D. Englund,
   \enquote{A quantum router architecture for high-fidelity entanglement flows in quantum networks,}
   {\protect\JournalTitle{npj Quantum Information}} \textbf{8}, 75 (2022).

\bibitem{repeater1}
  H.-J. Briegel, W. D\"ur, J. I. Cirac, and P. Zoller,
   \enquote{Quantum Repeaters: The Role of Imperfect Local Operations in Quantum Communication,}
   {\protect\JournalTitle{Physical Review Letters}} \textbf{81}, 5923--5935 (1998).

\bibitem{repeater2}
  N. Sangouard, C. Simon, H. Riedmatten, and N. Gisin,
   \enquote{Quantum repeaters based on atomic ensembles and linear optics,}
   {\protect\JournalTitle{Reviews of modern physics}} \textbf{83}, 33--80 (2011).

\bibitem{kanedaoptica}
  F. Kaneda, B. Christensen, J. Wong, H. Park, K. McCusker, and P. G. Kwiat,
   \enquote{Time-multiplexed heralded single-photon source,}
   {\protect\JournalTitle{Optica}} \textbf{2}, 1010--1013 (2015).

\bibitem{kanedaSA}
  F. Kaneda, and P. G. Kwiat,
   \enquote{High-efficiency single-photon generation via large-scale active time multiplexing,}
   {\protect\JournalTitle{Science Advances}} \textbf{5}, eaaw8586 (2019).
   
\bibitem{MccuskerPRL2009}
     K. T. McCusker, and P. G. Kwiat,
     \enquote{Efficient Optical Quantum State Engineering,}
      {\protect\JournalTitle{Physical Review Letters}} \textbf{103}, 163602 (2009).
      
\bibitem{oqc}
  J. L. O'Brien,
   \enquote{Optical Quantum Computing,}
   {\protect\JournalTitle{Science}} \textbf{318}, 1567--1570 (2007).
   
\bibitem{HOM_PRL1987}
     C.K. Hong, Z. Y. Ou, and L. Mandel,
     \enquote{Measurement of subpicosecond time intervals between two photons by interference,}
      {\protect\JournalTitle{Physical Review Letters}} \textbf{59}, 2044 (1987).

\bibitem{KLMNature2001}
     E. Knill, R. Laflamme, and G. J. Milburn,
     \enquote{A scheme for efficient quantum computation with linear optics,}
      {\protect\JournalTitle{Nature}} \textbf{409}, 46--52 (2001).
      
\bibitem{ultrafast}
  M. A. Hall, J. B. Altepeter, and P. Kumar,
   \enquote{Ultrafast Switching of Photonic Entanglement,}
   {\protect\JournalTitle{Physical Review Letters}} \textbf{106}, 053901 (2011).

\bibitem{THz}
  C. Kupchak, J. Erskine, D. England, and B. Sussman,
   \enquote{Terahertz-bandwidth switching of heralded single photons,}
   {\protect\JournalTitle{Optics Letters}} \textbf{44}, 1427--1430 (2019).

\bibitem{THzperspective}
  D. England, F. Bouchard, K. Fenwick, K. Bonsma-Fisher, Y. Zhang, P. J. Bustard, and B. J. Sussman,
   \enquote{Perspectives on all-optical Kerr switching for quantum optical applications,}
   {\protect\JournalTitle{Applied Physics Letters}} \textbf{119}, 160501 (2021).

\bibitem{cavity}
  K. T. McCusker, Y. Huang, A. S. Kowligy, and P. Kumar,
   \enquote{Experimental Demonstration of Interaction-Free All-Optical Switching via the Quantum Zeno Effect,}
   {\protect\JournalTitle{Physical Review Letters}} \textbf{110}, 240403 (2013).
   
\bibitem{waveguideEOM}
  V. Švarc, M. Nováková, G. Mazin, and M. Ježek,
   \enquote{Fully tunable and switchable coupler for photonic routing in quantum detection and modulation,}
   {\protect\JournalTitle{Optics Letters}} \textbf{44}, 5844--5847 (2019).

\bibitem{antonZ}
  X. Ma, S. Zotter, N. Tetik, A. Qarry, T. Jennewein, and A. Zeilinger,
   \enquote{A high-speed tunable beam splitter for feed-forward photonic quantum information processing,}
   {\protect\JournalTitle{Optics Express}} \textbf{19}, 22723--22730 (2011).

\bibitem{yariv}
  Y. Yariv and P. Yeh,
   \enquote{Photonics: Optical Electronics in Modern Communication (THE OXFORD SERIES IN ELECTRICAL AND COMPUTER ENGINEERING),}
   (Oxford University Press, Oxford, 2006) 6th ed., p.406
   
\bibitem{AOI1}
  A. Petrova-Mayor and S. Gimbal,
   \enquote{Advanced lab on Fresnel equations,}
   {\protect\JournalTitle{American Journal of Physics}} \textbf{83}, 935--941 (2015).

\bibitem{AOI2}
  A. Petrova-Mayor and S. Knudsen,
   \enquote{Analysis and manipulation of the induced changes in the state of polarization by mirror scanners,}
   {\protect\JournalTitle{Applied Optics}} \textbf{56}, 4513--4521 (2017).

\bibitem{JWP}
    H. Wang, J. Qin, X. Ding, M. Chen, S. Chen, X. You, Y. He, X. Jiang, L. You, Z. Wang, C. Schneider, Jelmer J. Renema, Sven Höfling, C. Lu, and J. Pan,
    \enquote{Boson Sampling with 20 Input Photons and a 60-Mode Interferometer in a \(10^{14}\)-Dimensional Hilbert Space,}
    {\protect\JournalTitle{Physical Review Letters}} \textbf{123}, 250503 (2019).

   \bibitem{silberhorn}
     E. Meyer-Scott, N. Prasannan, I. Dhand, C. Eigner, V. Quiring, S. Barkhofen, B. Brecht, M. B. Plenio, and C. Silberhorn,
     \enquote{Scalable Generation of Multiphoton Entangled States by Active Feed-Forward and Multiplexing,}
    {\protect\JournalTitle{Physical Review Letters}} \textbf{129}, 150501 (2022).

   \bibitem{xanadu}
     L. S. Madsen, F. Laudenbach, M. F. Askarani, F. Rortais, T. Vincent, J. F. F. Bulmer, F. M. Miatto, L. Neuhaus, L. G. Helt, M. J. Collins, A. E. Lita, T. Gerrits, S. W. Nam, V. D. Vaidya, M. Menotti, I. Dhand, Z. Vernon, N. Quesada and J. Lavoie,
     \enquote{Quantum computational advantage with a programmable photonic processor,}
     {\protect\JournalTitle{Nature}} \textbf{606}, 75--81 (2022).

   \bibitem{furusawa}
     J. Yoshikawa, K. Makino, S. Kurata, P. Loock, and A. Furusawa,
     \enquote{Creation, Storage, and On-Demand Release of Optical Quantum States with a Negative Wigner Function,}
     {\protect\JournalTitle{Physical Review X}} \textbf{3}, 041028 (2013).
   
\bibitem{process}
  Isaac L. Chuang and M. A. Nielsen
   \enquote{Prescription for experimental determination of the dynamics of a quantum black box,}
   {\protect\JournalTitle{Journal of Modern Optics}} \textbf{44}, 2455--2467 (1997).

\bibitem{tomography}
  D. F. V. James, P. G. Kwiat, W. J. Munro, and A. G. White,
   \enquote{Measurement of qubits,}
   {\protect\JournalTitle{Physical Review A}} \textbf{64}, 052312 (2001).

\bibitem{interface}
  J. Tang, Z. Hou, Q. Xu, G. Xiang, C. Li, and G. Guo,
   \enquote{Polarization-Independent Coherent Spatial-Temporal Interface with Low Loss,}
   {\protect\JournalTitle{Physical Review Applied}} \textbf{12}, 064058 (2019).

\bibitem{sagnac}
  A. Alarcón, P. González, J. Cariñe, G. Lima, and G. Xavier,
   \enquote{Polarization-independent single-photon switch based on a fiber-optical Sagnac interferometer for quantum communication networks,}
   {\protect\JournalTitle{Optics Express}} \textbf{28}, 33731--33738 (2020).

\bibitem{n_RTP}
   J. J. Carvajal, P. Segonds, A. Peña, J. Zaccaro, B. Boulanger, F. Díaz, and M. Aguiló,
    \enquote{Structural and optical properties of RbTiOPO4:Nb crystals,}
    {\protect\JournalTitle{Journal of Physics: Condensed Matter}} \textbf{19}, 116214 (2007).

\bibitem{migdall}
  A. L. Migdall, D. Branning, and S. Castelletto,
   \enquote{Tailoring single-photon and multiphoton probabilities of a single-photon on-demand source,}
   {\protect\JournalTitle{Physical Review A}} \textbf{66}, 053805 (2002).

\bibitem{pittman}
  T. B. Pittman, B. C. Jacobs, and J. D. Franson,
   \enquote{Single photons on pseudodemand from stored parametric down-conversion,}
   {\protect\JournalTitle{Physical Review A}} \textbf{66}, 042303 (2002).

\bibitem{multiphoton}
  J. Pan, Z. Chen, C. Lu, H. Weinfurter, A. Zeilinger, and M. \ifmmode \dot{Z}\else \.{Z}\fi{}ukowski,
   \enquote{Multiphoton entanglement and interferometry,}
   {\protect\JournalTitle{Review of Modern Physics}} \textbf{84}, 777 (2012).

\bibitem{multiplexedentangle}
  Z. Hou, J. Tang, C. Huang, Y. Huang, G. Xiang, C. Li, and G. Guo,
   \enquote{Entangled-State Time Multiplexing for Multiphoton Entanglement Generation,}
   {\protect\JournalTitle{Physical Review Applied}} \textbf{19}, L011002 (2023).

\bibitem{oneway}
  R.Raussendorf and H.J.Briegel,
   \enquote{A one-way quantum computer,}
   {\protect\JournalTitle{Physical Reveiw Letters}} \textbf{86}, 5188--5191 (2001).

\bibitem{metrology}
  B. L. Higgins, D. W. Berry, S. D. Bartlett, H. M. Wiseman, and G. J. Pryde,
   \enquote{Entanglement-free Heisenberg-limited phase estimation,}
   {\protect\JournalTitle{Nature}} \textbf{450}, 393--396 (2007).

 \end{thebibliography}


\end{document}